\documentclass{article}
\usepackage{spconf,amsmath,graphicx,cite}
\usepackage{multirow,amssymb}
\usepackage{xcolor,url}

\title{Multi-rate attention architecture for fast streamable Text-to-speech spectrum modeling}

\name{Qing He, Zhiping Xiu, Thilo Koehler, Jilong Wu\thanks{Correspondence to Qing He: qinghe@fb.com}}\address{Facebook AI}
	
\begin{document}

\maketitle

\begin{abstract}

Typical high quality text-to-speech (TTS) systems today use a two-stage architecture, with a spectrum model stage that generates spectral frames and a vocoder stage that generates the actual audio. High-quality spectrum models usually incorporate the encoder-decoder architecture with self-attention or bi-directional long short-term (BLSTM) units. While these models can produce high quality speech, they often incur O($L$) increase in both latency and real-time factor (RTF) with respect to input length $L$. In other words, longer inputs leads to longer delay and slower synthesis speed, limiting its use in real-time applications. In this paper, we propose a multi-rate attention architecture that breaks the latency and RTF bottlenecks by computing a compact representation during encoding and recurrently generating the attention vector in a streaming manner during decoding. The proposed architecture achieves high audio quality (MOS of 4.31 compared to groundtruth 4.48), low latency, and low RTF at the same time. Meanwhile, both latency and RTF of the proposed system stay constant regardless of input lengths, making it ideal for real-time applications.

\end{abstract}


\section{Introduction}
\label{sec:intro}

With the increasing popularity of voice assistant, virtual reality and other artificial intelligence technologies, text-to-speech (TTS) is becoming an important component in a wide range of applications. While recent advancements in neural TTS technologies have brought significant improvement in audio quality, efficient synthesis remains challenging in many scenarios \cite{wang2017tacotron, shen2018tacotron2, ren2019fastspeech, MSTransformerTTS, wavernn, oord2016wavenet,prenger2019waveglow,ping2017deep3}. In practical applications, latency, computational complexity, synthesis speed and streamability are key metrics for a production TTS system, especially if it has limited computational resources such as on mobile devices. 

TTS synthesis is a process of generating high-fidelity audio waveform using raw text representation. While solutions exists such as WaveNet\cite{oord2016wavenet} and WaveRNN\cite{wavernn} to generate speech waveform directly from text (with the help of a prosody model that predicts the tempo and intonation features), lower compute TTS such as Tacotron\cite{wang2017tacotron}, Tacotron2\cite{shen2018tacotron2}, FastSpeech\cite{ren2019fastspeech}, Deep voice 3\cite{ping2017deep3} etc, usually synthesize speech in two stages: 1: generate the speech spectrum from text; and 2: generate the speech waveform by conditioning on the predicted spectrum. We focus on the problem of two-stage TTS system design and propose a spectrum model that achieves low latency, supports streaming and produces high-quality TTS at the same time. 

Conventional high-quality spectrum models are usually based on the encoder-decoder attention framework \cite{wang2017tacotron, shen2018tacotron2, MSTransformerTTS}. In these models, the encoder summarizes the input utterance into a context representation and the decoder generates the spectral feature of each frame by attending to appropriate positions in the context. Since decoding starts after encoder computes the input utterance context, the latency of the model (i.e, time between compute starts and returning the first frame) is bounded by the compute time of the encoder, which increases proportionally with input length $L_\text{input}$ for both Bi-directional Long Short-term Memory (BLSTM) \cite{shen2018tacotron2} and self-attention \cite{ren2019fastspeech,MSTransformerTTS} models. Moreover, the size of the encoder output context is also proportional to the input length in these systems. Since decoding speed is affected by the size of the encoder context, the inference speed per frame becomes slower with increasing input length. Additionally, since the output spectrum is generated per frame, many systems use or upsample linguistic features to finer granularity for inference, which leads to increased synthesis computation complexity. 

In this paper, we propose a light-weight, low-latency and constant speed spectrum model. The model has the following characteristics: 1) the encoder adopts the parallelizable convolutional neural network (CNNs) and computes multi-head attention from linguistic source features at different levels (e.g., word, syllable, phone levels), leading to a compact context matrix and low-latency; 2) the model supports streaming as the multi-head attention vector is computed sequentially during decoding at each frame. 3) We perform dynamic pooling on the context matrix to restrict it to fixed length while still preserving information on the whole utterance. This ensures constant RTF during inference. Through MOS study and RTF measurements, we show that the model generates high quality speech while maintaining a constant (non-increasing) inference speed independent of input utterance lengths. 

\section{Related work}


Tacotron2 \cite{shen2018tacotron2} and FastSpeech \cite{ren2019fastspeech} are representative two-stage TTS systems that consist of an encoder-decoder spectrum model followed by a neural vocoder. Tacotron2's encoder incorporates layers of CNNs and a BLSTM, whereas the decoder computes an attention context from the encoder context and feeds it into a LSTM network with CNN post-net. The BLSTM in the encoder leads to $\mathbf{O}(L_\text{input})$ latency. Even though decoding is streamable, it's computation time per frame would be larger with longer utterances as it depends on the size of the encoder attention context. On the other hand, FastSpeech is a non-streaming model that generates the spectrum frames in parallel. Both of FastSpeech's encoder and decoder employ an architecture with layers of Feed-forward Transformer (FFT), which is composed of a multi-head self-attention layer and a 1D CNN layer. Even though parallel computation significantly reduces the total latency, the latency to return the first frame is still $\mathbf{O}(L_\text{frames})$ and the computation complexity is bounded by $\mathbf{O}(L_\text{frames}^2)$, which is not suitable for real-time streaming TTS applications on lower-resource devices. This is general for other transformer based models such as \cite{MSTransformerTTS}. 

Towards delivering a streaming solution with low-latency and constant inference speed regardless of input utterance length, we develop a multi-rate attention model architecture that utilizes a recurrent network with global attention context architecture similar to \cite{luong2015effective}. The global context is computed with multi-head attention as described in \cite{vaswani2017attention}. 

\section{Model Architecture}
\label{sec:model-arch}

\begin{figure}
  \centering
    \includegraphics[width=\linewidth]{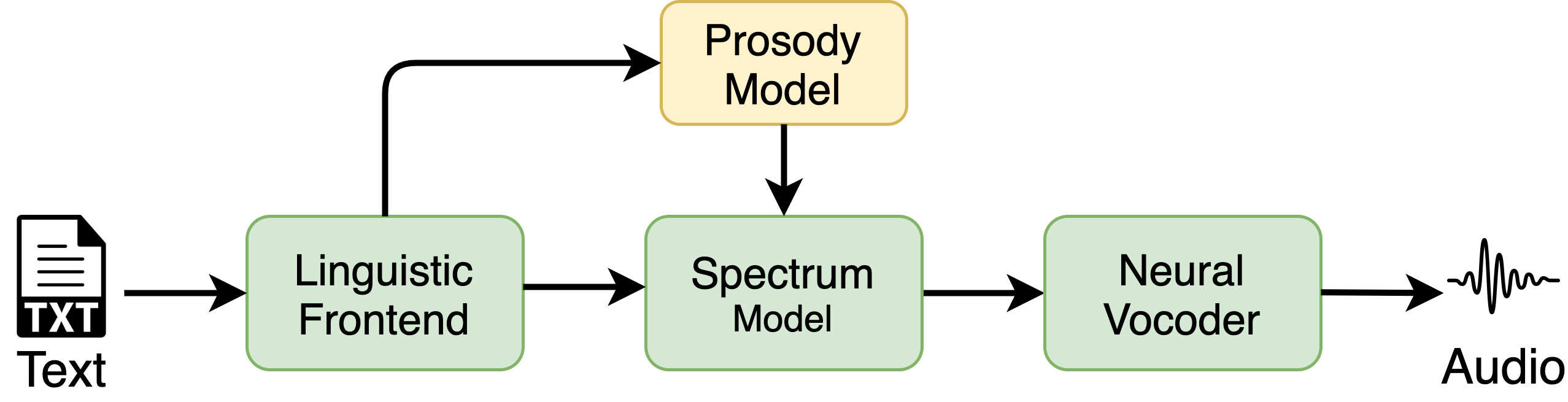}
    \caption{{\it Text to speech framework}: the linguistic frontend transforms text into an hierarchical IPA representation (Fig.\ref{fig:LCT}). The prosody model generates the phone durations. The spectrum model generates the spectrum frames which are fed into the neural vocoder to gernerate the audio waveforms.}
    \label{fig:Framework} 
\end{figure}

The TTS framework under investigation is shown in Fig.\ref{fig:Framework}. The multi-stage framework consists of a linguistic frontend that generates the linguistic representation of the input utterance at different levels (more details in Section \ref{sec:feature}); a prosody model that predicts the phone level duration, which is used to unroll the linguistic features; a spectrum model that takes the linguistic features to predict the spectrum frames; and a conditional neural vocoder that generates the audio waveform. Our proposed multi-rate attention architecture is a critical block in the spectrum model. 

\subsection{Multi-rate features}\label{sec:feature}

\begin{figure}[t]
  \centering
    \includegraphics[width=\linewidth]{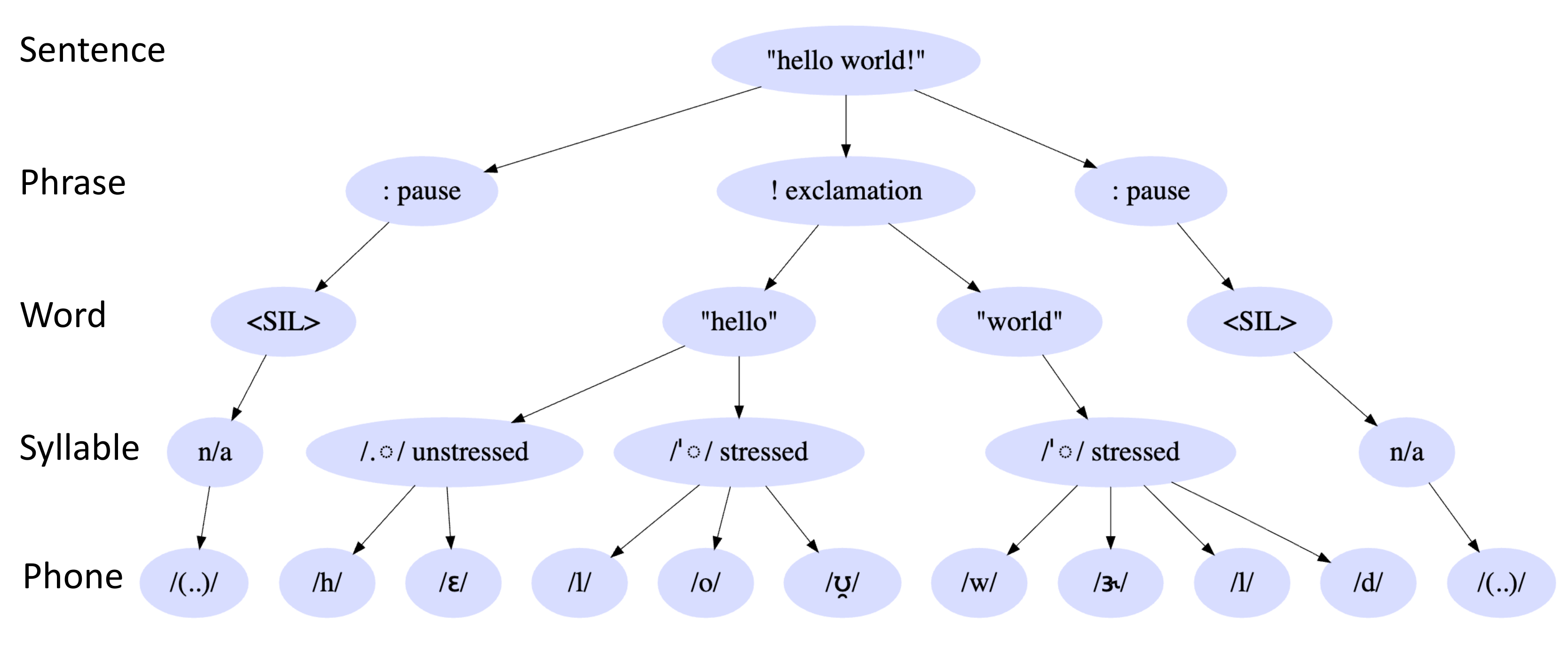}
    \caption{{\it{Linguistic context tree}} using the super-segmental IPA representation: the input text is transformed into a hierarchical representation of features containing information at different levels.}
    \label{fig:LCT} 
\end{figure}

TTS is a process of generating information and adding details. For an input utterance, there are different levels of information. Conventional TTS systems upsample lingustic information to a higher rate (e.g., sample rate \cite{oord2016wavenet}, frame rate \cite{ren2019fastspeech} and character rate \cite{shen2018tacotron2}). Instead, our system uses the features at their original information rates and perform computation on the low rate features. 

Specifically, our linguistic frontend adopts the super-segmental International Phonetic Alphabet (IPA) representation \cite{IPA}. As shown in Fig.\ref{fig:LCT}, the linguistic feature is an hierarchical representation of the normalized input text. There are sentence level features such as question/statement, language ID and speaker ID; phrase level features such as punctuation and intonation; word level features such as word embedding; syllable level features such as syllable type; and phone level features such as articulation diacritics\cite{IPA}. The linguistic features are rolled out using the phone level duration predictions from the prosody model to generate the frame level features, which contains information such as frame position in the current phone, phone duration and $\text{f}_0$, etc.  

\subsection{Multi-rate attention model}
\label{sec:multi-rateModel}

\begin{figure}[t]
  \centering
    \includegraphics[width=\linewidth]{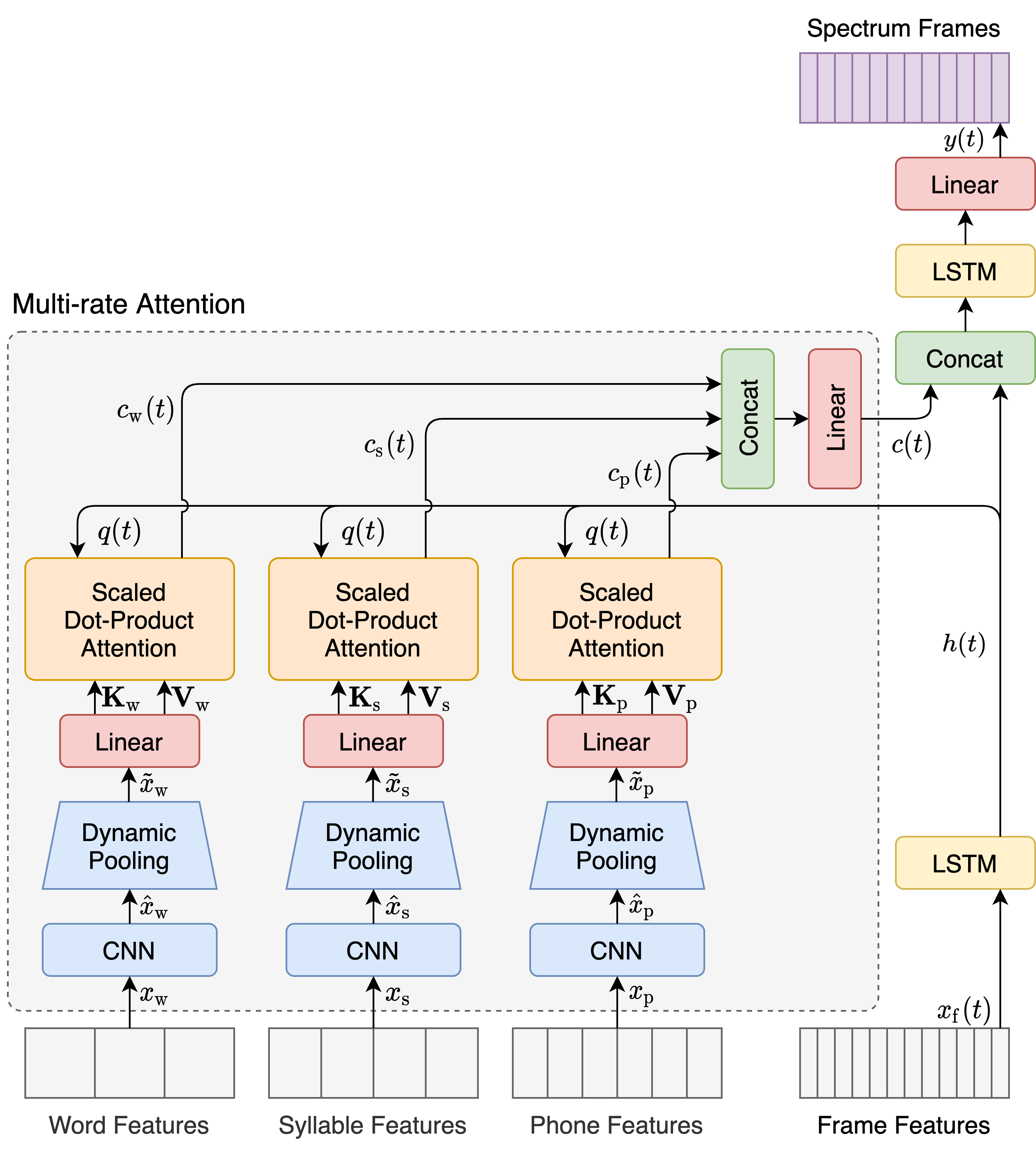}
    \caption{{\it Multi-rate attention architecture}: attention contexts are computed separately from each input source and then combined to provide overall context, $c(t)$, for each frame prediction. Each attention head contains compact linguistic information of the utterance at a difference IPA segmental level.}
    \label{fig:multi-rate} 
\end{figure}

The multi-rate recurrent attention architecture is shown in Fig.\ref{fig:multi-rate}. It adopts the global attention model as described in Section 3.1 of \cite{luong2015effective}. At each frame step $t$, the model first takes the corresponding frame level feature, $x_{\text{f}}(t)$, as input and runs it through one time-step in the LSTM. The hidden state $h(t)$ is used as the query vector to derive the context vector $c(t)$ that captures relevant information from each of the multi-rate source features, which are the word level feature, $x_\text{w}$, the syllable level feature, $x_\text{s}$ and the phone level feature, $x_\text{p}$. The input lengths of these features are $L_\text{w}$, $L_\text{s}$ and $L_\text{p}$, respectively. Then, the context $c(t)$ is concatenated with $h(t)$ to predict the current frame $y(t)$. Overall, the model is designed to predict the conditional distribution of the current frame given the current input and multi-rate linguistic context as follows: 
\begin{equation}
p\big(y(t)|y(<t),x_\text{f}(t),x_\text{w},x_\text{s},x_\text{p}\big).
\end{equation}
At each frame step, an attention vector, $c_\text{i}(t), \text{i} \in \{\text{w},\text{s},\text{p}\}$, is computed from each of the source input using the dot-product attention \cite{vaswani2017attention}: 
\begin{equation}
c_\text{i}\big(q(t),\mathbf{K}_\text{i},\mathbf{V}_\text{i}\big)(t) = \text{softmax}\left (\frac{q(t)\mathbf{K}_\text{i}^{T}}{\sqrt{d_{k_\text{i}}}}\right)\mathbf{V}_\text{i}, \hspace{0.2cm} \text{i} \in \{\text{w}, \text{s}, \text{p}\}
\end{equation}
where, the query $q(t) = h(t)$; keys, $\mathbf{K}_i$, and values, $\mathbf{V}_i$, are computed from the linguistic feature source, $\tilde{x}_{\text{i}}$, and have the same dimension $d_{k_\text{i}}$ and length $\tilde{L}_{\text{i}}$. The computation complexity of the attention vector $c_\text{i}(t)$ at each frame step is $\mathbf{O}(d_{k_\text{i}} \times \tilde{L}_{\text{i}})$, where $\tilde{L}_{\text{w}}$, $\tilde{L}_{\text{s}}$ and $\tilde{L}_{\text{p}}$ correspond to the lengths of $\tilde{x}_{\text{w}}$, $\tilde{x}_{\text{s}}$ and $\tilde{x}_{\text{p}}$, respectively, as shown in Fig.\ref{fig:multi-rate}. 

Finally, the overall context, $c(t)$ attends to information from different feature level and different positions using the multi-head mechanism as described in \cite{vaswani2017attention}. Specifically:
\begin{equation}
c(t) = \text{Concat}(c_\text{w},c_\text{s},c_\text{p})\mathbf{W}
\end{equation}
where $\mathbf{W}$ is a linear layer such that $\mathbf{W} \in \mathbb{R}^{(d_\text{w}+d_\text{s}+d_\text{p})\times d_\text{model}}$. 
  
\subsection{Context length regulation}
\label{sec:dynamicCNN}
As discussed in section \ref{sec:multi-rateModel}, the amount of compute for each attention vector is proportional to the size of the context matrix, i.e., $\mathbf{O}(d_{k_\text{i}} \times \tilde{L}_{\text{i}})$. We take two steps to reduce its complexity by reducing $\tilde{L}_{\text{i}}$. 

First, we use source information at its original compact representation in each hierarchical level without upsampling to a higher rate representation, which is a common practice in conventional TTS systems \cite{shen2018tacotron2,ren2019fastspeech,MSTransformerTTS}. The lengths $L_\text{w}$, $L_\text{s}$ and $L_\text{p}$ are orders of magnitude smaller than the model output length $L_\text{f}$ (i.e., number of spectrum frames). As a result, the overall computation is reduced by using the low-rate features. 

Secondly, in conventional systems, the length ${\tilde{L}}_{\text{i}}$ of the attention context matrix is equal to or proportional to the length of the input source ${L}_{\text{i}}, \text{i} \in \{\text{w}, \text{s}, \text{p}\}$ \cite{shen2018tacotron2,ren2019fastspeech,MSTransformerTTS}. This leads to increased computation complexity with increasing input utterance length. In order to maintain constant inference speed, we enforce a hard limit on the length of the attention context matrices (i.e, ${\tilde{L}}_{\text{i}}$) by performing dynamic max-pooling after the 1D CNN. This way, the reduced context matrix can still retain information of the entire input utterance. Dynamic max-pooling is done as follows:
\begin{equation}
    \hat{x}_\text{i}[m,n] = \sum^{d_\text{i}-1}_{c=0}\sum^{K-1}_{k=0} \mathbf{W}_\text{i}^m[c,k]x_\text{i}[n+k], \hspace{0.2cm} \text{i} \in \{\text{w}, \text{s}, \text{p}\}
\end{equation}
where, $0 \le m < d_{k_\text{i}}$, $0 \le n < L_\text{i}$, $d_\text{i}$ is the feature dimension of $x_\text{i}$, $K$ is CNN filter kernel size and $\mathbf{W}_\text{i}^m$ corresponds to the output filter matrix. Here, we assume the input is well padded and hence, $\hat{x}_\text{i}$ and $x_\text{i}$ have the same length $L_\text{i}$. 
After the CNN layer, we perform a dynamic max-pooling to limit the attention context length to be less or equal to a configurable value: $L_\text{i}^{\text{max}}$. Specifically,
\begin{equation}  
    \tilde{x}_\text{i}[m,n] = \max_{ n S_\text{i} \le \alpha \le n S_\text{i}+S_\text{i}-1} \hat{x}_\text{i}[m, \alpha], \hspace{0.2cm} \text{i} \in \{\text{w}, \text{s}, \text{p}\}
\end{equation}
where $0 \le m \le d_{k_\text{i}}$, $0 \le n < \tilde{L}_\text{i} = \min (L_\text{i},L_\text{i}^{\text{max}})$, and  $S_\text{i}$ is the max pooling stride such that $S_\text{i} = \lceil L_\text{i}/L_\text{i}^{\text{max}} \rceil$. Note that $\hat{x}_\text{i}$ is zero-padded to have length $S_\text{i} \times \tilde{L}_\text{i}$. 

\section{Experiment}
\subsection{Dataset and feature extraction}
The TTS dataset was recorded in a voice production studio by contracted professional voice talents. The corpus we used for training consists of 40,244 utterances from a single female speaker, which is approximately 40 hours of data with 24kHz sampling rate. The phone level duration feature is extracted with an unsupervised alignment algorithm, which uses the softDTW loss\cite{cuturi2017soft} and is trained using approximately $100$ hours of TTS multispeaker data. The $\text{f}_0$ feature is extracted for each frame using spectrum analysis. 

\subsection{Model details and baseline models}
The overall framework is described in Section \ref{sec:model-arch} and more details can be found in \cite{gao2020interactive}. 
The prosody model is a LSTM model with 256 hidden units with content-based global attention \cite{luong2015effective}. 
The spectrum feature is a 19-dim feature vector consisting of a 13-dim MFCC feature along with the $\text{f}_0$ feature and a 5-dim periodicity feature. In the multi-rate attention spectrum model, the CNN module has 2 layers of 1-D convolution with channel size set to be equal to the input feature dimensions; the dynamic pooling layer has a maximum sequence length, $L^{\text{max}}$, set to 50 for all heads; the dimension of the attention context $d_{k_i}$ is set to be equal to the dimension of the corresponding input features; the first and second layer of the LSTM has a hidden dimension of 256 and 128, respectively. The model is trained with the mean-square-error loss and the Adam \cite{adam} optimizer with learning-rate $1e^{-3}$ and decays with a factor of 0.85 for every 10 epochs. Lastly, the conditional neural vocoder is a WaveRNN \cite{wavernn} model, with hidden dimension 1024.  

We compare with several baseline models in terms of synthesis quality and inference speed. The inference speed metric is RTF, which is defined as (synthesis time)/(audio time). The LSTM baseline is a simple 2-layer LSTM model. The self-attention model uses frame rate features and the transformer architecture as described in \cite{vaswani2017attention}. We also compare with the Tacotron2 \cite{shen2018tacotron2} system, which uses the 80-dim mel spectrum features and the same WaveRNN vocoder \cite{shen2018tacotron2}.

\begin{table}[t]
\centering
\caption{MOS with 95\% confidence intervals}
\label{tab:MOS}
\begin{tabular}{|l|c|c|}
\hline
\textbf{Model}                                                                      & \textbf{Normal}         & \textbf{Long}  \\ \hline \hline
Groundtruth                                                                         & 4.48 $\pm$ 0.03                      & 4.58 $\pm$ 0.04                                      \\ \hline
Multi-rate attention (ours)                                                                & 4.31 $\pm$ 0.04                      & 4.08 $\pm$ 0.05                                     \\ \hline
\begin{tabular}[c]{@{}l@{}}Multi-rate attention\\ (no dynamic pooling)\end{tabular} & \multicolumn{1}{c|}{4.30 $\pm$ 0.04} & \multicolumn{1}{c|}{4.24 $\pm$ 0.05}                \\ \hline
LSTM                                                              & 3.72 $\pm$ 0.05                      & 3.85 $\pm$ 0.06                                     \\ \hline
Self-attention  \cite{vaswani2017attention}                                                                    & 4.21 $\pm$ 0.04                      & 4.05 $\pm$ 0.06                                     \\ \hline
Tacotron 2 \cite{Taco2_NV}                                                                         & 3.56 $\pm$ 0.05                      & 3.46 $\pm$ 0.08                                     \\ \hline
\end{tabular}
\vspace{-0.3cm}
\end{table}

\vspace{-0.1cm}
\subsection{Experiment results}
We have conducted two sets of mean opinion score (MOS) studies with testing utterances excluded from training data. Each MOS test has 400 participants who rate each sample between 1-5 (1:bad - 5:excellent). In the first test, we evaluate the general quality of the TTS using utterances with "normal" audio length ranging from 1 second to 20 seconds. In the second test, we focus on "long" utterances, with audio lengths ranging from 15 seconds to 40 seconds, in order to evaluate the model's ability to capture long range information and measure the effect of dynamic pooling when the stride is large. \footnote{Audio samples: https://multirate-spectrum-model.github.io/samples/} Table \ref{tab:MOS} shows the results of the MOS studies. Overall, the proposed model achieves better MOS scores than the baseline models. For normal utterances, the proposed multi-rate architecture delivers very natural TTS speech that is close to human speech. Dynamic pooling did not affect audio quality when utterances lengths are less than 20s. For long utterances, the gap between the proposed TTS voice and human voice increased because the prosody for longer utterances is harder to learn. In addition, we observed a small drop in audio quality when applying dynamic pooling with maximum context length $L^{max}=50$. 

\begin{figure}[t]
  \centering
    \includegraphics[width=0.9\linewidth, height=5.8cm]{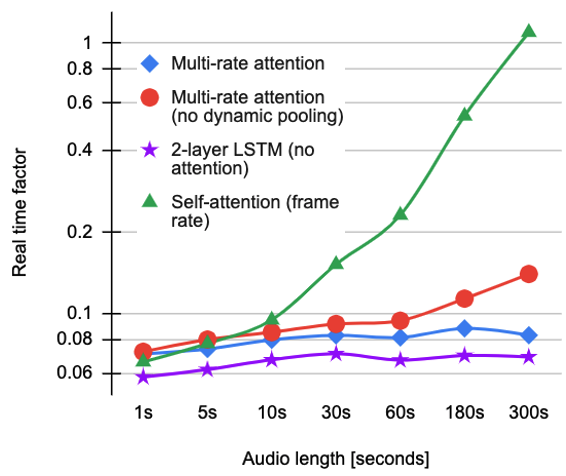}
        \caption{RTF evaluation with increasing audio length}
    \vspace{-0.4cm}
    \label{fig:RTF} 
\end{figure}

In addition, we evaluate the inference speed of the spectral model by measuring its RTF, which is an important metric for a production TTS. The RTF is measured as: (total synthesis time)/(audio length) using a single core on the Intel(R) Xeon(R) 2.0GHz CPU. As shown in Fig.\ref{fig:RTF}, the proposed multi-rate attention architecture achieves almost constant RTF with increasing input utterance length, whereas the RTF of the baseline self-attention model increases proportionally with audio length. Moreover, the multi-rate attention model RTF is only slightly higher than the simple 2-layer LSTM model. Notice that, when we remove dynamic pooling from the model, the RTF starts to increase when input length is long. Hence, there is a small trade-off between RTF and audio quality with long utterances and maximum context length, $L^{max}$, can be tuned accordingly based on the quality and speed requirement of the target application.

\section{Conclusion}
In conclusion, we proposed a spectrum model using the multi-rate attention architecture. The proposed model uses the compact hierarchical linguistic features as input, computes the multi-head attention contexts in a streaming manner and performs dynamic pooling to restrict context length. It achieves constant latency and RTF regardless of input utterance lengths while producing high quality speech. Our MOS study shows the proposed model achieves better quality compared to the self-attention baseline, as well as other baseline models. 






\vfill\pagebreak

\bibliographystyle{IEEEbib}
\bibliography{refs}

\end{document}